\documentclass[twocolumn]{svjour3}
\usepackage{natbib}
\setcitestyle{round}
\usepackage{graphicx}
\usepackage{subfigure,graphicx}
\usepackage{amsmath}
\usepackage[]{placeins}
\usepackage{amsfonts}
\usepackage{float}
\usepackage{amssymb}
\usepackage{mathtools}
\usepackage{tensor}
\usepackage{enumitem}
\usepackage{xcolor}
\def\bfs{{\bf s}}

\def\bfx{{\bf x}}
\def\bfy{{\bf y}}

\def\bfC{{\bf C}}

\def\bfN{{\bf N}}

\def\bfR{{\bf R}}
\def\bfS{{\bf S}}

\def\bfX{{\bf X}}
\def\bfF{{\bf F}}
\def\bfU{{\bf U}}

\def\bfe{{\bf e}}

\def\l{\lambda}

\def\e0{\varepsilon_0}

\newcommand\sts{\sigma_{\texttt{ts}}}

\newcommand\shs{\sigma_{\texttt{hs}}}

\renewcommand\e{\varepsilon}

\long\def\symbolfootnote[#1]#2{\begingroup%
\def\thefootnote{\fnsymbol{footnote}}\footnote[#1]{#2}\endgroup}

\begin{document}

\journalname{International Journal of Fracture}
\titlerunning{Validating Griffith fracture propagation in the phase-field approach to fracture}

\title{\Large{Validating Griffith fracture propagation in the phase-field approach to fracture: The case of Mode III by means of the trousers test}}

\author{F. Kamarei \and E. Breedlove \and  O. Lopez-Pamies}

\institute{
           F. Kamarei \at Department of Civil and Environmental Engineering, University of Illinois, Urbana--Champaign, IL 61801-2352, USA\\
           \email{kamarei2@illinois.edu}\vspace{0.1cm}
           \and
           E. Breedlove \at 3M Corporate Research Laboratory,  St. Paul, MN 55144, USA \\
           \email{elbreedlove@mmm.com}\vspace{0.1cm}
           \and
           Oscar Lopez-Pamies \at Department of Civil and Environmental Engineering, University of Illinois, Urbana--Champaign, IL 61801-2352, USA \\
           \email{pamies@illinois.edu}
           }

\date{}
\maketitle

\begin{abstract}

At present, there is an abundance of results showing that the phase-field approach to fracture in elastic brittle materials --- when properly accounting for material strength --- describes the \emph{nucleation} of fracture from large pre-existing cracks in a manner that is consistent with the Griffith competition between bulk deformation energy and surface fracture energy. By contrast, results that demonstrate the ability of this approach to describe Griffith fracture \emph{propagation} are scarce and restricted to Mode I. Aimed at addressing this lacuna, the main objective of this paper is to show that the phase-field approach to fracture describes Mode III fracture propagation in a manner that is indeed consistent with the Griffith energy competition. This is accomplished via direct comparisons between phase-field predictions for fracture propagation in the so--called \emph{trousers} \emph{test} and the corresponding results that emerge from the Griffith energy competition. The latter are generated from full-field finite-element solutions that --- as an additional critical contribution of this paper --- also serve to bring to light the hitherto unexplored limitations of the classical Rivlin-Thomas-Greensmith formulas that are routinely used to analyze the trousers test.

\keywords{Toughness; Flow and Fracture; Energy Methods; Computational Mechanics}

\end{abstract}

\vspace{0.5cm}

\section{Introduction}\label{Sec: Intro}

Since a first version was introduced at the turn of the millennium \citep{Bourdin00}, well understood and correctly implemented phase-field models of fracture for elastic brittle materials that properly account for material strength have been repeatedly shown to describe the \emph{nucleation} of fracture from large pre-existing cracks in a manner that is consistent with the Griffith competition between bulk deformation energy and surface fracture energy; see, e.g., \cite{Muller10,McMeeking15,Landis17,Tanne18,KBFLP20,Kumar25}. By contrast, studies that demonstrate the ability of phase-field models to describe Griffith fracture \emph{propagation} are scarce and, what is more, restricted to Mode I\footnote{In this work, we employ the general material-independent definition that Mode I, Mode II, and Mode III refer, respectively, to the opening, sliding, and tearing modes of crack growth.}; see, e.g., \cite{Tanne18,KBFLP20,KLP20}. Presumably, this is because while there is a multitude of boundary-value problems for which Griffith fracture nucleation from large pre-existing cracks is available in closed form and hence allow for a facile direct comparison \citep{Tada73}, there are only a few known boundary-value problems for which Griffith fracture propagation occurs in a stable and analytically tractable manner. In particular, as far as we know,  there is the family of untapered and tapered double cantilever beam tests for Mode I fracture propagation \citep{Gilman1960,Gilman64,Moses1968,Ripling67}. For Mode III fracture propagation, there is the trousers test \citep{RT53,Greensmith55,SR74}.

In this context, the main objective of this paper is to show that the phase-field approach to fracture describes Mode III fracture propagation in a manner that is indeed consistent with the Griffith energy competition. We do so by directly comparing phase-field predictions for the fracture propagation in the trousers test with the corresponding results that emerge from the Griffith energy competition. The latter are generated from full-field finite-element (FE) solutions, which --- as an additional critical contribution of this paper --- permit to also examine, apparently for the first time, the limitations of the classical formulas worked out by \cite{RT53} and \cite{Greensmith55} in their pioneering studies of the trousers test. 

We begin in Section \ref{Sec: Trousers} by reviewing the trousers test, recalling the classical formulas routinely used to describe how fracture propagates in such a test according to the Griffith energy competition (Subsections \ref{Sec: GRT} and \ref{Sec: GTadot}), and probing the accuracy of these formulas via direct comparisons with full-field FE solutions for an elementary case, that of a trousers test for a specimen made of a Neo-Hookean material (Subsection \ref{FEvsGRT}). In Section \ref{Sec: Phase-field Theory}, we briefly summarize the phase-field model introduced by \cite*{KFLP18}. In contrast to the original version \citep{Bourdin00} and subsequent modifications with energy splits (see, e.g., the reviews of energy splits included in \cite{Wick22,Perego24}), this phase-field model accounts not only for the elasticity and the toughness of the material, but also for its strength as an independent macroscopic material property, and it is for this reason that it has the ability to describe nucleation of fracture at large, in addition to describing fracture propagation; see \cite{KLP20,KBFLP20,LPDFL25,KDLP25}. Finally, in Section \ref{Sec: Simulations Non-Linear}, we make use of the general phase-field model presented in Section \ref{Sec: Phase-field Theory} to simulate the same canonical trousers test examined in Subsection \ref{FEvsGRT} in order to probe its description of Mode III fracture propagation vis-\`a-vis that of the Griffith energy competition.
%We close by recording a number of final comments in Section \ref{Sec: Final comments}. 

\section{The trousers test}\label{Sec: Trousers}

The trousers test is one of three famous tests that \cite{RT53} introduced to investigate the validity of the Griffith energy criterion
\begin{equation}
G:=-\dfrac{\partial \mathcal{W}}{\partial \mathrm{\Gamma}}=G_c \label{Gc-0}
\end{equation}
to describe the \emph{nucleation} of fracture from a large crack in elastic brittle materials within the setting of finite quasi-static deformations. In expression (\ref{Gc-0}), the left-hand side $-\partial \mathcal{W}/\partial \mathrm{\Gamma}$ denotes the change in stored elastic energy in the specimen at hand with respect to an added surface area ${\rm d}\mathrm{\Gamma}$ to an existing crack $\mathrm{\Gamma}$ in its reference state --- under conditions of fixed deformation of those parts of the boundary that are not traction-free so that no work is done by the external forces --- while the right-hand side $G_c$ stands for the toughness or critical energy release rate, a macroscopic material property. In a companion contribution, \cite{Greensmith55} recognized that the trousers test is also particulary expedient to investigate the validity of the Griffith energy criterion (\ref{Gc-0}) to describe fracture \emph{propagation}. As elaborated in the sequel, this is because the crack in such a test can be made to propagate at a constant speed, $\dot{a}$ say, that is directly controlled by the applied boundary conditions. 

\begin{figure}[b!]\centering
 \includegraphics[width=3.2in]{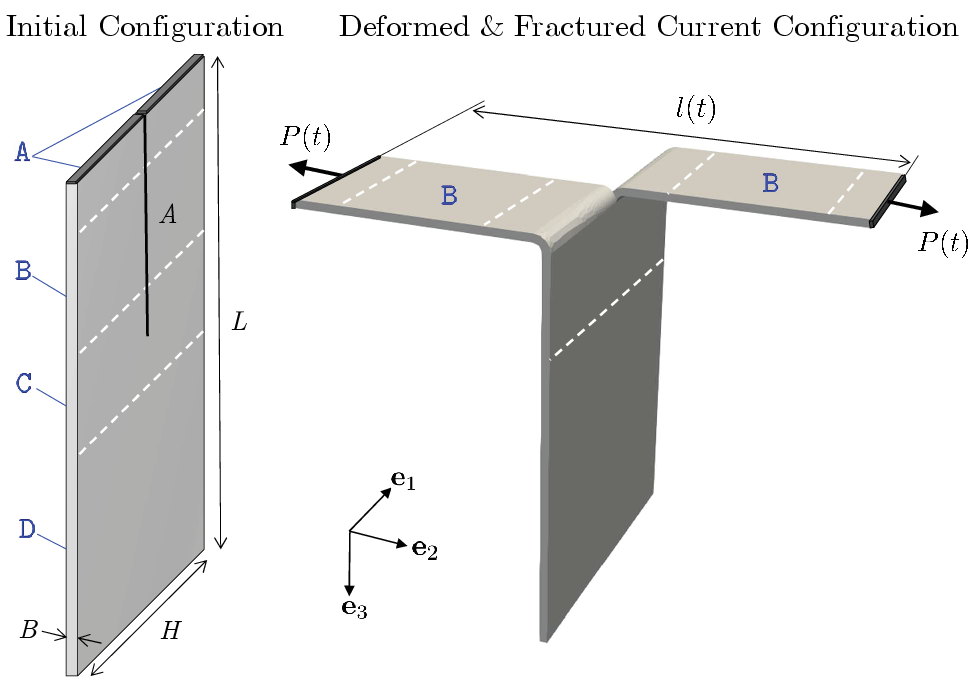}
\caption{\small Schematic of the trousers test. The dimensions in the initial configuration are such that $B\ll H<A < L$. The two bottom ends of the trousers legs are held firmly by stiff grips and then pulled apart either by an applied force $P(t)$ or by an applied deformation $l(t)$.}
   \label{Fig1}
\end{figure}
Figure \ref{Fig1} provides a schematic of a typical trousers test. Distinctly, in its initial configuration, the thickness of the specimen is much smaller than its height, $B\ll H$, its height is smaller than its length, $H<L$, and the length of the pre-existing crack is larger than the height of the specimen but smaller than its length, $H< A< L$.

The specimen is loaded by first bending its two legs in opposite directions to bring them into the same plane. Subsequently, their bottom ends are held firmly by stiff grips, which are then pulled apart; think of the tearing of a paper sheet ($\sim0.1$ mm in thickness) or an aluminum foil ($\sim0.01$ mm in thickness). The pulling is done by either applying a force $P(t)$ or a deformation $l(t)$ over a time interval $t\in[0,T]$. When a force is applied, it is typically ramped up over an initial period $[0,t_0]$  and then held constant. When a deformation is applied, on the other hand, the grips are typically separated at a constant rate $\dot{l}_0$, resulting in a current grip-to-grip separation of $l(t)=l_0+\dot{l}_0 t$. 

\subsection{The classical approximation $G_{RT}$ of Rivlin and Thomas (1953) for the energy release rate $G$}\label{Sec: GRT}

Consider a trousers test in which the specimen is of a suitable size\footnote{Specimens with either too large or too small legs, relative to the elasticity and the toughness of the material that they are made of, may \emph{not} deform and fracture in a manner that is self-similar.} that it deforms and fractures in a self-similar manner. Granted this self-similarity, at any given time during the test, as illustrated by Fig.~\ref{Fig1}, there are four different regions of deformation: $\texttt{A}$, $\texttt{B}$, $\texttt{C}$, and $\texttt{D}$.  Region $\texttt{D}$ is substantially undeformed, the crack-front region $\texttt{C}$ and the grip region $\texttt{A}$ are in a complex state of deformation (highly non-uniform in space), while region $\texttt{B}$ is expected to be roughly in a state of spatially uniform uniaxial tension.

Now, at any given time $t\in(0,T]$ during the test, when the length of the crack is $a(t)$ and the force at the grips is $P(t)$, imagine an increase in the crack surface of amount ${\rm d}\mathrm{\Gamma}=B {\rm d}a$. This increase in crack surface does not affect the complex deformation states in regions $\texttt{A}$ and $\texttt{C}$, nor their sizes, but rather translates region $\texttt{C}$ along the direction of the added crack resulting in the growth of region $\texttt{B}$ at the expense of region $\texttt{D}$. Precisely, an added crack ${\rm d}\mathrm{\Gamma}$ at constant force $P(t)$ would result in the transferring of a volume $H{\rm d}\mathrm{\Gamma}$ of the specimen from an undeformed state in region $\texttt{D}$ to a state of approximate uniaxial tension in region $\texttt{B}$. Making use of this observation, we can immediately deduce the estimate
\begin{equation}
\left.\dfrac{\partial \mathcal{W}}{\partial \mathrm{\Gamma}}\right|_{P}=H W_{\texttt{ut}}(\lambda(t)),\label{ERR-P}
\end{equation}
where the suffix $P$ denotes differentiation at fixed force $P(t)$, $\lambda(t)$ is the stretch that results by subjecting the material to uniform uniaxial nominal stress $2P(t)/(BH)$, and $W_{\texttt{ut}}$ stands for the elastic energy density stored at that state of deformation.

By definition, the derivative in the Griffith energy criterion (\ref{Gc-0}) for a trousers test is to be taken at fixed deformation $l(t)$, and \emph{not} at fixed force $P(t)$. To determine $\partial \mathcal{W}/\partial \mathrm{\Gamma}|_{l}$, with some abuse of notation, we write
\begin{equation*}
l(t)=l(P(t),\mathrm{\Gamma})
\end{equation*}
so that the deformation $l(t)$ between the grips can be viewed as implicitly characterized in terms of the force $P(t)$ at the grips and the size $\mathrm{\Gamma}$ of the crack, and note that 
\begin{align*}
{\rm d} \mathcal{W}=&\left.\dfrac{\partial \mathcal{W}}{\partial l}\right|_{\mathrm{\Gamma}}{\rm d}l+\left.\dfrac{\partial \mathcal{W}}{\partial \mathrm{\Gamma}}\right|_{l}{\rm d} \mathrm{\Gamma}=P(t){\rm d}l+\left.\dfrac{\partial \mathcal{W}}{\partial \mathrm{\Gamma}}\right|_{l}{\rm d} \mathrm{\Gamma},
\end{align*}
where we have made use of the connection $P(t)=\partial\mathcal{W}/\partial l|_{\mathrm{\Gamma}}$. Upon using the condition ${\rm d}P=0$, it follows that
\begin{align*}
\left.-\dfrac{\partial \mathcal{W}}{\partial \mathrm{\Gamma}}\right|_{l}=& P(t)\dfrac{{\rm d}l}{{\rm d}\mathrm{\Gamma}}-\dfrac{{\rm d}\mathcal{W}}{{\rm d} \mathrm{\Gamma}}=P(t)\left.\dfrac{\partial l}{\partial \mathrm{\Gamma}}\right|_{P}-\left.\dfrac{\partial \mathcal{W}}{\partial \mathrm{\Gamma}}\right|_{P}.
\end{align*}
Making use of the estimate
\begin{equation}
\left.\dfrac{\partial l}{\partial a}\right|_{P}=2\lambda(t),\quad \text{ so that } \quad\left.\dfrac{\partial l}{\partial \mathrm{\Gamma}}\right|_{P}=\dfrac{2}{B}\lambda(t),\label{dellP}
\end{equation}
and invoking the result (\ref{ERR-P}), it finally follows that
\begin{equation}
-\dfrac{\partial \mathcal{W}}{\partial \mathrm{\Gamma}}=\dfrac{2}{B} P(t)\lambda(t)-H W_{\texttt{ut}}(\lambda(t))=:G_{RT},\label{Gc-Trousers}
\end{equation}
where we have reverted back to omitting the suffix $l$ in the derivative $-\partial \mathcal{W}/$ $\partial \mathrm{\Gamma}$, since there is no longer risk of confusion, and have labeled the resulting energy release rate $G_{RT}$, after Rivlin and Thomas. Observe that:
\begin{enumerate}[label=\textit{\roman*.}]

\item{As expected \citep{Griffith21}, the formula (\ref{Gc-Trousers}) is nothing more than the change in the work done by the applied forces, $2P(t)\lambda(t) /B$, minus the change in stored elastic energy, $H W_{\texttt{ut}}(\lambda(t))$, per surface ${\rm d}\mathrm{\Gamma}$ of added crack, or, in other words, the negative of the change in potential energy per surface of added crack;}
     
\item{Irrespective of whether the test is carried out by prescribing a force $P(t)$ or a deformation $l(t)$, the computation of the formula (\ref{Gc-Trousers}) amounts simply to having knowledge of the response of the material under uniaxial tension, via $\lambda(t)$ and $W_{\texttt{ut}}(\lambda(t))$, and of the force $P(t)$ at the grips;}
    
\item{By construction, the formula (\ref{Gc-Trousers}) applies exclusively to specimens that deform and fracture in a self-similar manner, for which the state of deformation in region $\texttt{B}$ is one of uniform uniaxial tension, and for which the relation between crack growth and grip separation is described by (\ref{dellP});}
    
\item{For linear elastic brittle materials, $\lambda(t)=1$, $W_{\texttt{ut}}=2P^2(t)/(B^2H^2E)$, and the formula (\ref{Gc-Trousers}) specializes to
\begin{equation*}
-\dfrac{\partial \mathcal{W}}{\partial \mathrm{\Gamma}}=\dfrac{2}{B} P(t)-\dfrac{2}{B^2 H E}P^2(t),%\label{Gc-Trousers-Linear}
\end{equation*}
where $E$ stands for the Young's modulus of the material.
    }
    
\item{For sufficiently thin specimens, $\lambda(t)\approx1$, the contribution $H W_{\texttt{ut}}(\lambda(t))$ is negligible, and the formula (\ref{Gc-Trousers}) renders the remarkably simple approximation
\begin{equation*}
-\dfrac{\partial \mathcal{W}}{\partial \mathrm{\Gamma}}=\dfrac{2}{B} P(t).%\label{Gc-Trousers-Limit}
\end{equation*}
}

\end{enumerate}

\subsection{The classical results of Greensmith and Thomas (1955) for fracture propagation under constant grip separation rate}\label{Sec: GTadot}

Granted the result (\ref{Gc-Trousers}), it follows from the Griffith energy criterion (\ref{Gc-0}) that crack propagation in a trousers test takes place whenever the condition
\begin{equation}
\dfrac{2}{B} P(t)\lambda(t)-H W_{\texttt{ut}}(\lambda(t))=G_c\label{Griffith-Trousers-Cond} 
\end{equation}
is satisfied. For the case when the deformation $l(t)$ is applied at a constant rate $\dot{l}_0$, the force at the grips increases monotonically in time from $P(0)=0$ at $t=0$ until some critical time $t_c$ at which the value $P_c\equiv P(t_c)$ satisfies the condition (\ref{Griffith-Trousers-Cond}) and the crack starts to grow. Assuming that the process of crack growth is continuous in time, after remaining put at $a(t)=A$ for $0\leq t< t_c$, the crack length grows according to
\begin{align}
a(t)=&A+\dfrac{\dot{l}_0}{2\lambda_c} (t-t_c)\nonumber\\
=&A-\dfrac{l_0+\dot{l}_0 t_c}{2\lambda_c}+\dfrac{1}{2\lambda_c}l(t)\label{Crack-Growth-Trousers}
\end{align}
for $t\geq t_c$. Here, $\lambda_c\equiv\lambda(t_c)$ is the stretch associated with the critical value $P_c$ of the force $P(t)$, which remains at the constant value $P(t)=P_c$ for $t\geq t_c$ during the entire crack propagation process, as a result of the criticality condition (\ref{Griffith-Trousers-Cond}) being independent of the crack length. Observe that:
\begin{enumerate}[label=\textit{\roman*.}]

\item{The speed of crack propagation 
\begin{equation}
\dot{a}(t)=\dfrac{\dot{l}_0}{2\lambda_c}\label{Crack-Growth-Speed-Trousers}
\end{equation}
is constant and given directly in terms of the rate $\dot{l}_0$ at which the grips are separated. 
}

\item{In stark contrast to the speed of crack propagation (\ref{Crack-Growth-Speed-Trousers}), the critical value $P_c$ of the force at which the crack propagates is independent of how fast the grips are separated.
}
     
\item{For linear elastic brittle materials, the critical value $P_c$ of the force at which the crack propagates is given by
\begin{align*}
P_c=&\dfrac{1}{2}\left(E-\sqrt{E^2-\dfrac{2 E G_c}{H}}\right)B H.%\label{Pc-Linear}
\end{align*}
 }
 
\item{For sufficiently thin specimens, when $\lambda_c\approx 1$ and the contribution $H W_{\texttt{ut}}(\lambda_c)$ is negligible, the remarkably simple approximations
\begin{align*}
P_c=\dfrac{B G_c}{2}\quad {\rm and}\quad \dot{a}(t)=\dfrac{\dot{l}_0}{2}%\label{Pc-a-Thin}
\end{align*}
apply.
 }
    
\end{enumerate}

\subsection{Full-field FE results for the energy release rate $G$ versus the Rivlin-Thomas approximation $G_{RT}$}\label{FEvsGRT}

Notwithstanding the elegance and simplicity of the preceding \emph{approximate} analysis, given any material of interest, it is \emph{not} immediately apparent how to choose appropriate dimensions for a trousers specimen so that the formula (\ref{Gc-Trousers}) --- and hence (\ref{Griffith-Trousers-Cond}) and (\ref{Crack-Growth-Trousers}) --- applies to that material. 

In the literature, as well as in the \cite{ASTM}, following  \cite{RT53} and \cite{Greensmith55}, specimens of length $L\in[100,150]$ mm, height $H\in[10,40]$ mm, thickness $B\in[1,2]$ mm, and initial crack length $A=50$ mm are believed to be of appropriate dimensions for the formula (\ref{Gc-Trousers}) to apply, at least when dealing with elastomers. In this subsection, we show that such a belief is misplaced. We do so by comparing the formula (\ref{Gc-Trousers}) directly with results for $G$ obtained from full-field FE solutions for an elementary example, that of a trousers test for a specimen made of a Neo-Hookean material. 

\subsubsection{The specimen and its reference configuration}

For definiteness, we consider a specimen of dimensions identical to those used by \cite{Greensmith55} in their pioneering trousers experiments on natural rubber and SBR, to wit, we consider a specimen of length $L=100$ mm and height $H=40$ mm in the $\bfe_3$ and $\bfe_1$ directions and constant thickness $B=1$ mm in the $\bfe_2$ direction. The specimen contains a pre-existing edge crack of length $A=50$ mm in the $\bfe_3$ direction. Here, $\{\bfe_i\}$ stands for the laboratory frame of reference. Its origin is place at the center of the crack front; see Fig.~\ref{Fig1}. 

As noted above, a trousers specimen is loaded by first bending its two legs in opposite directions to bring them into the same plane, at which point their bottom ends are firmly gripped and pulled apart in the loading frame. For this reason, it proves convenient not to use the initial configuration as the reference configuration to carry out the simulations, but to use, instead, the configuration as initially mounted in the loading frame. In such a configuration, following in the footsteps of \cite{SLP23c}, we take the transition from the legs to the un-cracked part of the specimen to be circular fillets of inner radius $R_f$ plus an additional straight segment of length $A_b$, so that the reference configuration of the specimen is given by $\mathrm{\Omega}_0=\mathrm{\Omega}_0^{(\mathcal{L})}\bigcup\mathrm{\Omega}_0^{(\mathcal{L}_f)}\bigcup\mathrm{\Omega}_0^{(\mathcal{L}_b)}\bigcup\mathrm{\Omega}_0^{(\mathcal{R}_b)}$ $\bigcup\mathrm{\Omega}_0^{(\mathcal{R}_f)}\bigcup\mathrm{\Omega}_0^{(\mathcal{R})}$ $\bigcup\mathrm{\Omega}_0^{(\mathcal{B})}$, where
$\mathrm{\Omega}_0^{(\mathcal{L})}=\{\bfX:-H/2<X_1<0,\,-l_0/2<X_2<-B/2-R_f,\,-A_b-R_f-B<X_3<-A_b-R_f\}$, $\mathrm{\Omega}_0^{(\mathcal{L}_f)}=\{\bfX:-H/2<X_1<0,\,-B/2-R_f<X_2,\,-A_b-R_f-B<X_3,\,R^2_f<(X_2+B/2+R_f)^2+(X_3+A_b)^2<(R_f+B)^2\}$,
$\mathrm{\Omega}_0^{(\mathcal{L}_b)}=\{\bfX:-H/2<X_1<0,\,-B/2<X_2<B/2,\,-A_b<X_3<0\}$, 
$\mathrm{\Omega}_0^{(\mathcal{R}_b)}=\{\bfX:0<X_1<H/2,\,-B/2<X_2<B/2,\,-A_b<X_3<0\}$, $\mathrm{\Omega}_0^{(\mathcal{R}_f)}=\{\bfX:0<X_1<H/2,\,X_2<B/2+R_f,\,-R_f-B<X_3,\,R^2_f<(X_2-B/2-R_f)^2+(X_3-A_b)^2<(R_f+B)^2\}$, $\mathrm{\Omega}_0^{(\mathcal{R})}=\{\bfX:0<X_1<H/2,\,B/2+R_f<X_2<l_0/2,\,-A_b-R_f-B<X_3<-A_b-R_f\}$, and 
$\mathrm{\Omega}_0^{(\mathcal{B})}=\{\bfX:-H/2<X_1<H/2,\,-B/2<X_2<B/2,\,0<X_3<L-A\}$. Note that in this reference configuration the initial grip separation is $l(0)=l_0=2(A-A_b)+B-(\pi-2)R_f$.

Numerical experiments show that the reference configuration $\mathrm{\Omega}_0$ spelled out above can be idealized to be undeformed and stress free without having a significant impact on the response of the specimen, provided that the separation between the grips $l(t)$ is sufficiently larger than the initial separation $l_0$; typically, $l(t)>2A+B$ suffices. The same numerical experiments also show little dependence on the specific values of the radius $R_f$ of the fillet and the length $A_b$ of the additional straight segment. The simulations presented below correspond to $R_f=5B=5$ mm and $l_0=91.29$ mm.

\subsubsection{The elasticity of the material}

As announced from the outset, to keep the results as fundamental as possible, we consider that the specimen is made of a homogeneous material with compressible Neo-Hookean elasticity. Precisely, its stored-energy function is given by
\begin{equation}
W(\bfF)=\dfrac{\mu}{2}\left(I_1-3\right)-\mu\ln J+\dfrac{\mathrm{\Lambda}}{2}(J-1)^2,\label{W-NH}
\end{equation}
where $\bfF$ is the deformation gradient tensor, $I_1=\bfF\cdot\bfF={\rm tr}\, \bfC$, $J=\det\bfF=\sqrt{\det\bfC}$ stand for the first and third principal invariants of the right Cauchy-Green deformation tensor $\bfC$, and where $\mu$ and $\mathrm{\Lambda}$ are material constants. It follows that the first Piola-Kirchhoff stress at any material point $\bfX\in\mathrm{\Omega}_0$ and time $t\in[0,T]$ is given by
\begin{equation*}
\bfS(\bfX,t)=\dfrac{\partial W}{\partial \bfF}(\bfF)=\mu(\bfF-\bfF^{-T})+\mathrm{\Lambda}(J-1)J\bfF^{-T}.%\label{S-NH}
\end{equation*}

The simulations presented below correspond to the values $\mu=0.5$ MPa and $\mathrm{\Lambda}=1$ MPa, the first of which is representative of the initial shear moduli of the SBR elastomers studied by \cite{Greensmith55}. The value of $\mathrm{\Lambda}=1$ MPa, which is smaller than those typical of standard elastomers such as SBR, is chosen here merely for convenience, so that the resulting governing equations can be solved with standard FEs. In any case, the value of $\mathrm{\Lambda}$ has no fundamental impact on the results \citep{KBLP25}.

\subsubsection{The FE computation of $G$}

\begin{figure}[b!]\centering
 \includegraphics[width=3.2in]{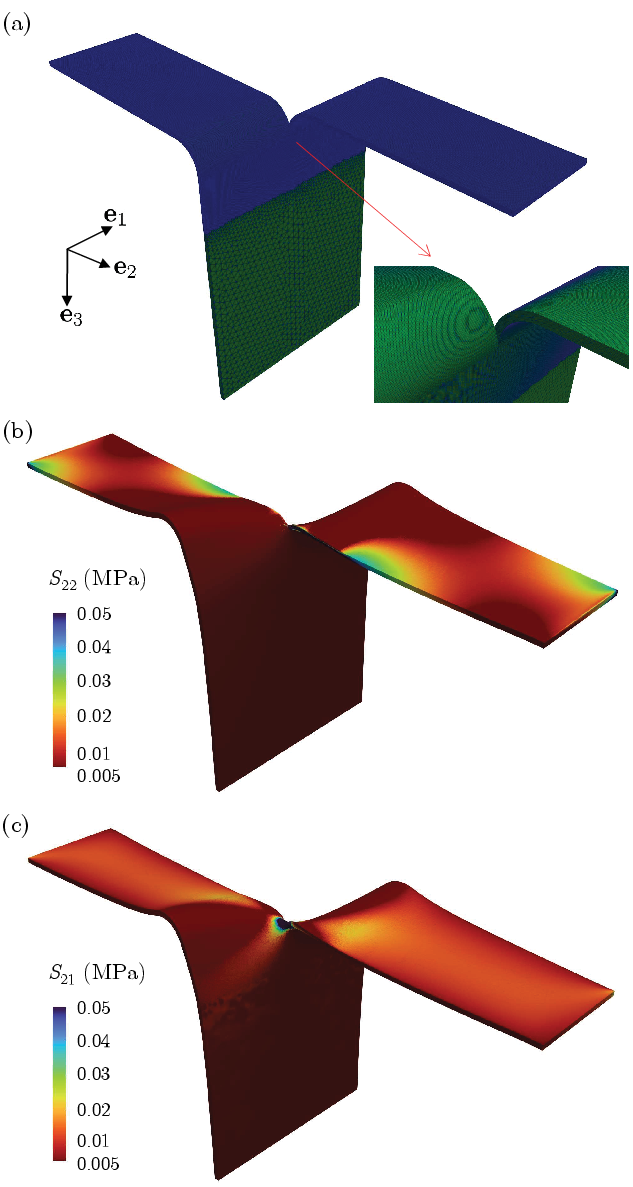}
\caption{\small (a) One of the FE meshes used to compute the energy release rate (\ref{FE-G}) for the Neo-Hookean trousers tests. (b,c) Contour plots of the stress components $S_{22}(\bfX,t)$ and $S_{21}(\bfX,t)$ over the deformed configuration of the specimen shown in part (a) at the applied separation $l(t)=106$ mm between the grips. The figures pertain to the specimen with crack length $A=50$ mm.}
   \label{Fig2}
\end{figure}

In order to compute the energy release rate $G$ from full-field results, in addition to the specimen with crack length $A=A_0=50$ mm, we consider eight additional specimens with larger crack lengths $A=A_r=A_0+r\Delta A$, for $r=1,...,8$ with $\Delta A=1$ mm, solve by means of the FE method the elastostatic responses of all nine of them for a range of applied deformations $l(t)\in[l_0,l_f]=[91.29,115]$ mm, compute their respective total elastic energies $\mathcal{W}(l,A)$ $=\int_{\mathrm{\Omega}_0}W(\bfF)\,{\rm d}\bfX$, and finally use the finite-difference quotient
\begin{equation}
G=-\dfrac{\mathcal{W}(l,A_r+\Delta A)-\mathcal{W}(l,A_r)}{\Delta A},\label{FE-G}
\end{equation}
$r=0,...,8$.

We remark that the crack increment $\Delta A=1$ mm is small enough to lead to accurate quotients (\ref{FE-G}). 

We also remark that a mesh of linear simplicial elements of size $\texttt{h}=0.125$ mm in the legs and around the crack suffices to generate converged solutions. Part (a) of Fig.~\ref{Fig2} presents one such mesh for the specimen with crack length $A=50$ mm. For completeness, parts (b) and (c) of the figure show the contour plots of the stress components $S_{22}(\bfX,t)$ and $S_{21}(\bfX,t)$ over the deformed configuration of the specimen at the separation $l(t)=106$ mm between the grips. These latter parts make it clear that the legs of the specimen undergo significant bending and twisting, in addition to stretching, and that the classical approximation that the region $\texttt{B}$ is in a state of uniform uniaxial tension is \emph{not} justified.

\subsubsection{Results}

\begin{figure}[b!]\centering
 \includegraphics[width=2.8in]{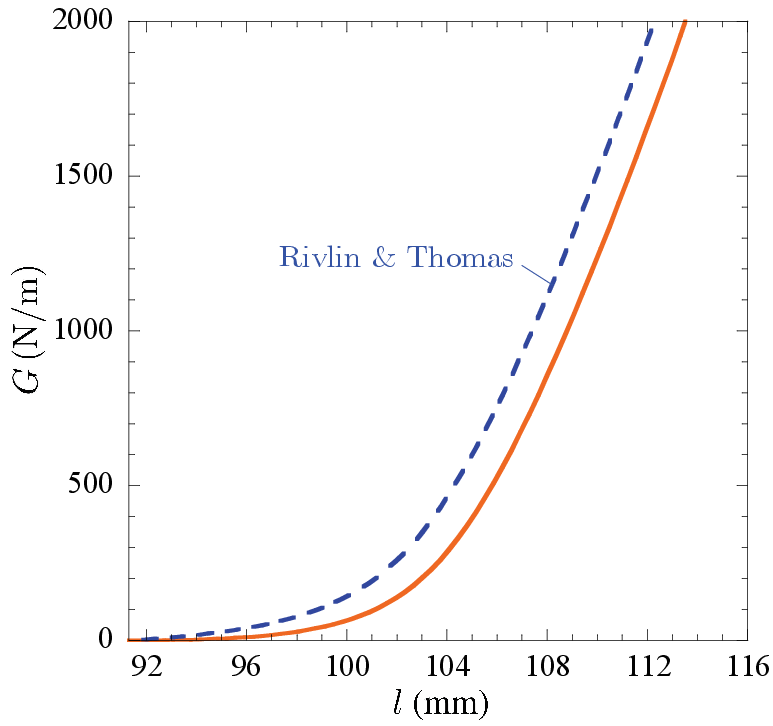}
\caption{\small Comparison between the energy release rate $G$ computed from full-field FE results (solid line) and the Rivlin-Thomas approximation $G_{RT}$ for the Neo-Hookean trousers test with crack length $A=50$ mm. The results are shown as a function of the separation $l(t)$ between the grips.}
   \label{Fig3}
\end{figure}
\begin{figure}[t!]\centering
 \includegraphics[width=2.8in]{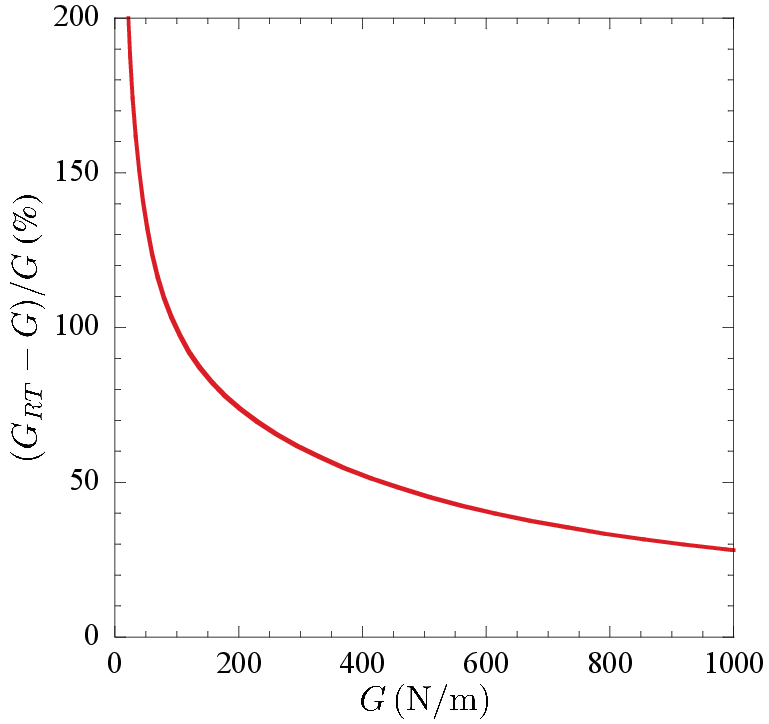}
\caption{\small Relative error $(G_{RT}-G)/G$ of the Rivlin-Thomas approximation $G_{RT}$ with respect to the energy release rate $G$ computed from full-field FE results for the Neo-Hookean trousers test with crack length $A=50$ mm. The result is shown as a function of $G$.}
   \label{Fig4}
\end{figure}

Figure \ref{Fig3} compares the energy release rate (\ref{FE-G}) determined from full-field FE solutions and the classical approximation (\ref{Gc-Trousers}) from \cite{RT53} for the Neo-Hookean trousers test with crack length $A=50$ mm. The results are shown as a function of the applied deformation $l(t)$ between the grips. A quick glance suffices to recognize that while the Rivlin-Thomas approximation $G_{RT}$ is qualitatively similar to the FE result for $G$, it significantly overpredicts it. Figure \ref{Fig4} provides a plot of the relative error $(G_{RT}-G)/G$ as a function of the value of $G$.

%
%%%%%%%%%%%%%%%%%%%%%%%%%%%%%%%%%%%%%%%%%%%%%%%%%%%%%%%%%%%%%%%%%%%%%%%%%%%%%%
\begin{figure}[t!]
  \subfigure[]{
   \label{fig:4a}
   \begin{minipage}[]{0.5\textwidth}
   \centering \includegraphics[width=2.8in]{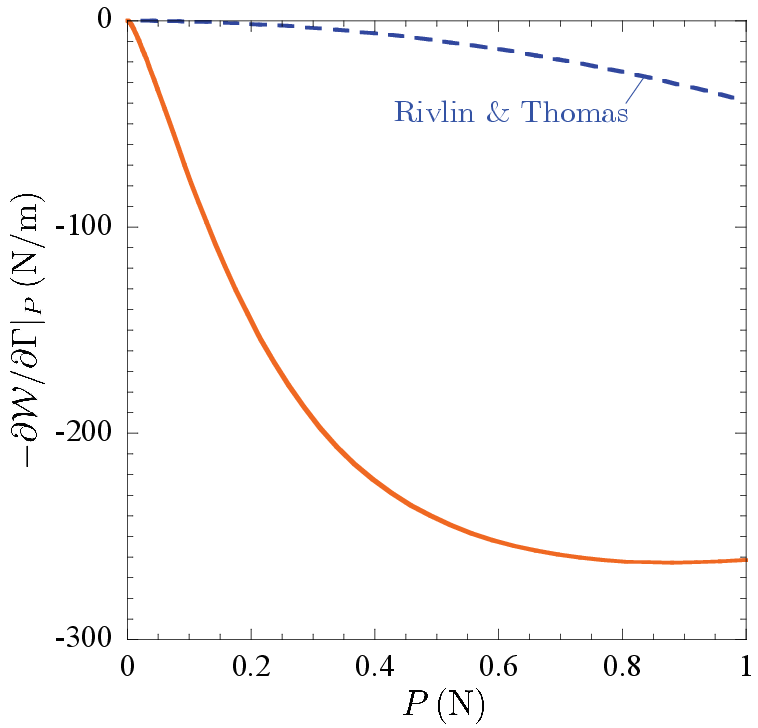}
   \vspace{0.1cm}
   \end{minipage}}
  \subfigure[]{
   \label{fig:4b}
   \begin{minipage}[]{0.5\textwidth}
   \centering \includegraphics[width=2.8in]{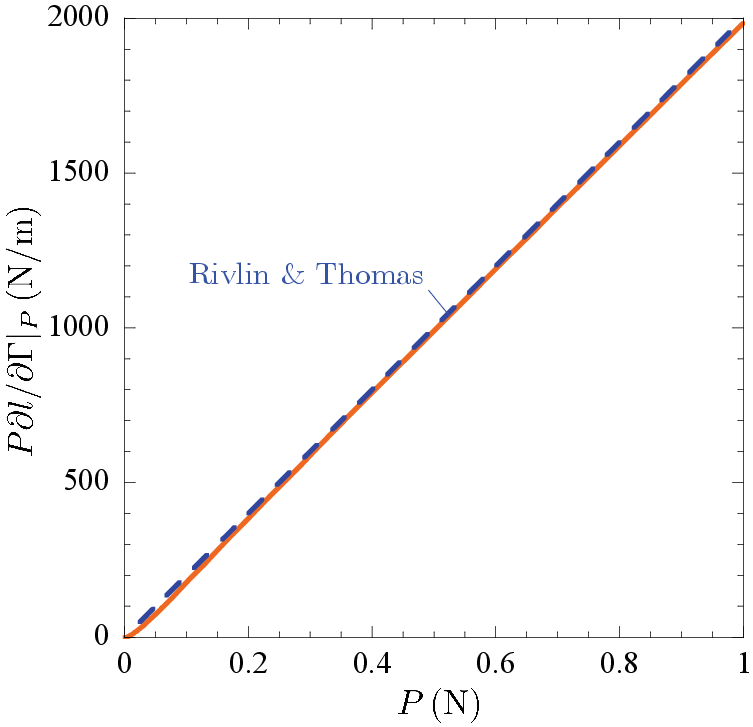}
   \vspace{0.1cm}
   \end{minipage}}
   \caption{Comparisons between the two different contributions in the energy release rate $G=P(t)\partial l/\partial \mathrm{\Gamma}|_{P}-\partial \mathcal{W}/\partial \mathrm{\Gamma}|_{P}$ computed from full-field FE results (solid lines) and from the Rivlin-Thomas approximations for the Neo-Hookean trousers test with crack length $A=50$ mm. (a) The change in stored elastic energy $-\partial \mathcal{W}/\partial \mathrm{\Gamma}|_{P}$. (b) The change in the work done by the external forces $P(t)\partial l/\partial \mathrm{\Gamma}|_{P}$. The results are shown as a function of the force $P(t)$ at the grips.}\label{Fib5}
\end{figure}
%%%%%%%%%%%%%%%%%%%%%%%%%%%%%%%%%%%%%%%%%%%%%%%%%%%%%%%%%%%%%%%%%%%%%%%%%%%%%%
%

To pinpoint the source of the discrepancy between $G_{RT}$ and the actual $G$, Fig.~\ref{Fib5} plots separately the two different contributions to $G$, namely, the change in stored elastic energy $-\partial \mathcal{W}/\partial \mathrm{\Gamma}|_{P}$ and the change in the work done by the external forces $P(t)\partial l/\partial \mathrm{\Gamma}|_{P}$, as determined from the full-field FE solutions and from the Rivlin-Thomas approximations (\ref{ERR-P}) and (\ref{dellP}). The results are plotted as a function of the force $P(t)$ at the grips.

It is immediate from Fig.~\ref{Fib5} that the Rivlin-Thomas approximation $G_{RT}$ is inaccurate primarily because its estimate (\ref{ERR-P}) for the change in stored elastic energy $-\partial \mathcal{W}/\partial \mathrm{\Gamma}|_{P}$ is inaccurate. In particular, the estimate (\ref{ERR-P}) underpredicts the actual result because it does not account for the bending and twisting of the legs of the specimen. By contrast, its estimate (\ref{dellP}) for the relation between crack growth and grip separation, and hence for the change in the work done by the external forces $P(t)\partial l/\partial \mathrm{\Gamma}|_{P}$, does provide a good approximation.

\begin{remark}
\emph{The comparisons in Figs. \ref{Fig3}, \ref{Fig4}, and \ref{Fib5} make it plain that the Rivlin-Thomas formula (\ref{Gc-Trousers}) can produce substantially inaccurate results when applied to trousers tests that use specimens of standard dimensions. As a corollary of significant practical implications, they indicate that numerous trousers experiments that have been analyzed over the years with such a formula to extract the fracture toughness $G_c$ of elastomers would need to be reanalyzed.   }
\end{remark}

\begin{figure}[t!]\centering
 \includegraphics[width=2.8in]{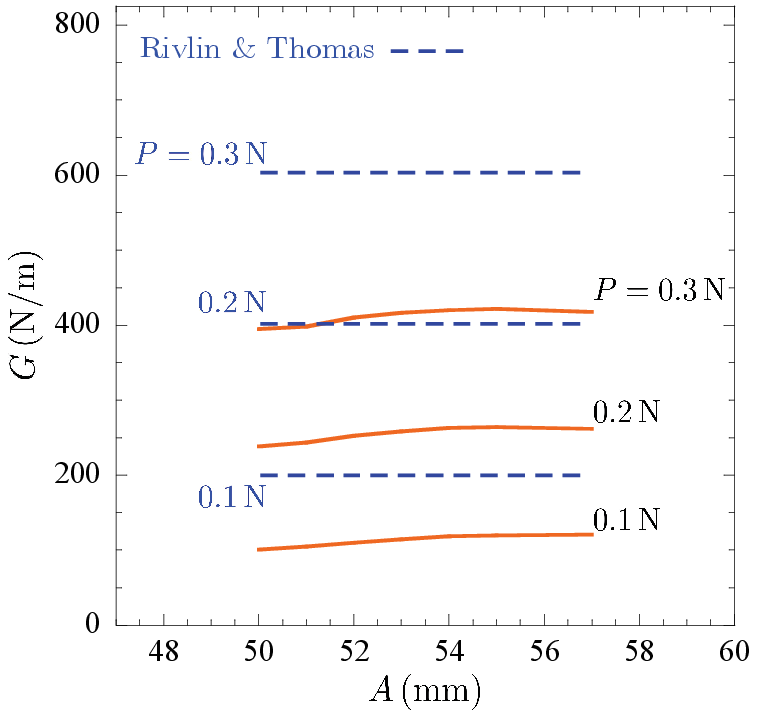}
\caption{\small Comparison between the energy release rate $G$ computed from full-field FE results (solid lines) and the Rivlin-Thomas approximation $G_{RT}$ for the Neo-Hookean trousers test. The results are shown as a function of the crack length $A$ for three different constant values of the force $P(t)$ at the grips.}
   \label{Fig6}
\end{figure}

We close this subsection by presenting in Fig.~\ref{Fig6} results for the energy release rate $G$ as a function of crack length $A$ for three different constant values of the force at the grips, $P(t)=0.1, 0.2, 0.3$ N. Again, the results from the Rivlin-Thomas approximation are included for direct comparison with the those obtained from the full-field FE solutions. 

The main observation from Fig.~\ref{Fig6} is that $G$ depends very weakly on the length of the crack for fixed values of the force $P(t)$ at the grips, which implies that crack propagation is grosso modo stable at constant applied force, as observed in experiments \citep{Greensmith55}. Moreover, as already noted above, the Rivlin-Thomas approximation $G_{RT}$ is \emph{not} just weakly but exactly independent of crack length and, again, clearly quantitatively inaccurate.

\section{The phase-field approach to fracture}\label{Sec: Phase-field Theory}

Next, we briefly recall the phase-field formulation introduced by \cite*{KFLP18} to describe the nucleation and propagation of fracture in elastic brittle materials subjected to quasi-static mechanical loading. For simplicity of exposition and later comparison with the preceding results, we present the model for the basic case of elastic brittle materials with Neo-Hookean elasticity. For a complete account, including FE schemes to solve the resulting governing equations, the interested reader is referred to \cite{KRLP18,KLP20,KBFLP20,KRLP22,KKLP24}.

\subsection{Kinematics and boundary conditions}

Consider a specimen that, initially, at time $t=0$, occupies the open bounded domain $\mathrm{\Omega}_0$. We denote its boundary by $\partial \mathrm{\Omega}_0$, its outward unit normal by $\bfN$, and identify material points by their initial position vector $\bfX\in \mathrm{\Omega}_0$. The specimen is subjected to a deformation $\overline{\bfy}(\bfX,t)$ on a part $\partial \mathrm{\Omega}_0^{\mathcal{D}}$ of the boundary, and a surface force (per unit undeformed area) $\overline{\bfs}(\bfX,t)$ on the complementary part $\partial \mathrm{\Omega}_0^{\mathcal{N}}=\partial\mathrm{\Omega}_0\setminus\partial \mathrm{\Omega}_0^{\mathcal{D}}$; for simplicity, body forces are considered to be absent. In response to these boundary conditions --- all of which are assumed to be applied monotonically and quasi-statically in time --- the position vector $\bfX$ of a material point in the specimen will move to a new position specified by 
\begin{equation}
\bfx=\bfy(\bfX,t),\label{def-mapping}
\end{equation}
where $\bfy(\bfX,t)$ is the deformation field. We write the associated deformation gradient at $\bfX$ and $t$ as $\bfF(\bfX,t)=\nabla\bfy(\bfX,t)$. In addition to the deformation (\ref{def-mapping}), the boundary conditions may result in the nucleation and subsequent propagation of cracks in the specimen. We describe such cracks in a regularized fashion via the phase field
\begin{equation*}
z=z(\bfX,t)
\end{equation*}
taking values in the range $[0, 1]$. The value $z=1$ identifies the intact regions of the material and $z=0$ the regions that have been fractured, while the transition from $z=1$ to $z=0$ is set to occur smoothly over regions of small thickness of regularization length scale $\varepsilon>0$.

\subsection{Constitutive behavior}

The specimen is taken to be made of a homogeneous, isotropic, elastic brittle material. Its mechanical behavior is hence characterized by three material properties: its elasticity, its strength, and its toughness. 

As noted above, we take the elastic behavior of the material to be characterized by the Neo-Hookean stored-energy function (\ref{W-NH}). 

The strength of the material is taken to be characterized by the Drucker-Prager strength surface
\begin{equation}
\mathcal{F}(\bfS)=\sqrt{\mathcal{J}_2}+\dfrac{s_{\texttt{ts}}}
{\sqrt{3}\left(3 s_{\texttt{hs}}-s_{\texttt{ts}}\right)} \mathcal{I}_1-\dfrac{\sqrt{3}s_{\texttt{hs}} s_{\texttt{ts}}}
{3s_{\texttt{hs}}-s_{\texttt{ts}}}=0,\label{DP}
\end{equation}
where $\mathcal{I}_1=s_1+s_2+s_3$ and $\mathcal{J}_2=1/3(s_1^2+s_2^2+s_3^2-s_1 s_2-s_1 s_3-s_2s_3)$ stand for invariants of the Biot stress tensor $\bfS^{(1)}=(\bfS^T\bfR+\bfR^T\bfS)/2$, with $s_1$, $s_2$, $s_3$ denoting the principal Biot stresses and $\bfR$ the rigid rotation from the polar decomposition $\bfF=\bfR\bfU$, while the constants $s_{\texttt{ts}}>0$ and $s_{\texttt{hs}}>0$ stand for the uniaxial tensile and hydrostatic strengths of the material. That is, they denote the critical nominal stress values at which fracture nucleates under uniform states of monotonically increased uniaxial tension $\bfS={\rm diag}(s>0,0,0)$ and hydrostatic stress $\bfS={\rm diag}(s>0,s>0,s>0)$, respectively.

Finally, we take the critical energy release rate of the material to be given by the constant
$$G_c.$$

\subsection{Governing equations}

According to the phase-field fracture formulation put forth by \citet*{KFLP18}, the deformation field $\bfy_k(\bfX)=\bfy(\bfX,t_k)$ and phase field $z_k(\bfX)=z(\bfX,t_k)$ at any material point $\bfX \in \overline{\mathrm{\Omega}}_0=\mathrm{\Omega}_0\cup\partial\mathrm{\Omega}_0$ and at any given discrete time $\{t_{k}\}_{k=0,1,\ldots,M}$, with $t_{0} = 0$ and $t_{M} = T$, are determined by the system of coupled partial differential equations 

\begin{equation}
\left\{\hspace{-0.05cm}\begin{array}{ll}
{\rm Div}\left[z_{k}^2 \left(2\mu(\nabla \bfy_{k}-\nabla \bfy_{k}^{-T})+\right.\right. &\vspace{0.1cm}\\
\hspace{0.7cm}\left.\mathrm{\Lambda}(J_k-1)J_k\nabla \bfy_{k}^{-T})\right]=\textbf{0},& \,\bfX\in\mathrm{\Omega}_0\vspace{0.2cm}\\
\bfy_k(\bfX)=\overline{\bfy}(\bfX,t_k), & \; \bfX\in\partial\mathrm{\Omega}_0^{\mathcal{D}}\vspace{0.2cm}\\
z_{k}^2 \left(2\mu(\nabla \bfy_{k}-\nabla \bfy_{k}^{-T})+\right. &\vspace{0.1cm}\\
\mathrm{\Lambda}(J_k-1)J_k\nabla \bfy_{k}^{-T})\bfN=\overline{\bfs}(\bfX,t_k),& \; \bfX\in\partial\mathrm{\Omega}_0^{\mathcal{N}}
\end{array}\right. \label{BVP-y-theory}
\end{equation}
and
\begin{equation}
\left\{\hspace{-0.05cm}\begin{array}{ll}
 \varepsilon\, \delta^\varepsilon G_c \triangle z_k=\dfrac{8}{3}z_{k} W(\nabla\bfy_k)-\dfrac{4}{3}c_\texttt{e}(\bfX,t_{k})-&\vspace{0.1cm}\\
\hspace{1.8cm}\dfrac{\delta^\varepsilon G_c}{2\varepsilon}+\dfrac{8}{3\,\zeta} \, p(z_{k-1},z_k),& \bfX\in \mathrm{\Omega}_0
\vspace{0.2cm}\\
\nabla z_k\cdot\bfN=0,&  \bfX\in \partial\mathrm{\Omega}_0
\end{array}\right. \label{BVP-z-theory}
\end{equation}
with $p(z_{k-1},z_k)=|z_{k-1}-z_k|-(z_{k-1}-z_k)-|z_k|+z_k$. In these equations, $\nabla \bfy_k(\bfX)=\nabla \bfy(\bfX,t_k)$, $J_k=\det\nabla\bfy_k$, $\nabla z_k(\bfX)=\nabla z(\bfX,t_k)$, $\triangle z_k(\bfX)=\triangle z(\bfX,t_k)$, we recall that the stored-energy function $W(\bfF)$ is given by expression (\ref{W-NH}), $\zeta$ is a penalty parameter\footnote{The penalty function $p(z_{k-1},z_k)$ and penalty parameter $\zeta$ in (\ref{BVP-z-theory}) enforce that the phase field remains in the physically admissible range $0\leq z\leq 1$ and that fracture is irreversible. These requirements can be enforced by means of different strategies \citep{Wick15}, the penalty approach spelled out here being one of them.} such that $\zeta^{-1}\gg \delta^\varepsilon G_c/(2\varepsilon)$, and, making use of the constitutive prescription introduced in \citep{KKLP24}, 
\begin{align*}
c_{\texttt{e}}(\bfX,t)&=z^2\beta_2\sqrt{\mathcal{J}_2}+z^2\beta_1\mathcal{I}_1+z\left(1-\dfrac{\sqrt{\mathcal{I}^2_1}}{\mathcal{I}_1}\right)W(\bfF)%\label{ce}
\end{align*}
with 

\begin{align*}
&\beta_1=-\dfrac{1}{\shs}\delta^\varepsilon\dfrac{G_c}{8\varepsilon}+\dfrac{2W_{\texttt{hs}}}{3\shs},\\
&\beta_2=-\dfrac{\sqrt{3}(3\shs-\sts)}{\shs\sts}\delta^\varepsilon\dfrac{G_c}{8\varepsilon}-
\dfrac{2W_{\texttt{hs}}}{\sqrt{3}\shs}+\dfrac{2\sqrt{3}W_{\texttt{ts}}}{\sts},
\end{align*}
and
\begin{equation*}
\delta^\varepsilon=\left(\dfrac{\sts+(1+2\sqrt{3})\,\shs}{(8+3\sqrt{3})\,\shs}\right)\dfrac{3 G_c}{16W_{\texttt{ts}}\varepsilon}+\dfrac{2}{5}.
%\label{delta-eps-final}
\end{equation*}
In these last expressions, $W_{\texttt{ts}}$ and $W_{\texttt{hs}}$ stand for the values of the stored-energy function (\ref{W-NH}) along uniform uniaxial tension and hydrostatic stress states at which the strength surface (\ref{DP}) is violated. To wit,
\begin{equation*}
W_{\texttt{ts}}=\dfrac{\mu}{2}(\l^2_{\texttt{ts}}+2\l^2_l-3)-\mu\ln(\l_{\texttt{ts}}\l_l^2)+\dfrac{\mathrm{\Lambda}}{2}(\l_{\texttt{ts}}\l_l^2-1)^2%\label{Wt}
\end{equation*}
and
\begin{equation*}
W_{\texttt{hs}}=\dfrac{\mu}{2}(3\l^2_{\texttt{hs}}-3)-\mu\ln(\l^3_{\texttt{hs}})+\dfrac{\mathrm{\Lambda}}{2}(\l^3_{\texttt{hs}}-1)^2,\label{WtsWhs}
\end{equation*}
where the pair of stretches ($\l_{\texttt{ts}}, \l_l$) and the stretch $\l_{\texttt{hs}}$ are defined implicitly as the roots closest to $(1,1)$ and $1$ of the system of nonlinear algebraic equations
\begin{align*}
&\left\{\begin{array}{l}
s_{\texttt{ts}}=\mu(\l_{\texttt{ts}}-\l_{\texttt{ts}}^{-1})+\mathrm{\Lambda}(\l_{\texttt{ts}}\l_l^4-\l_l^2)\\[10pt]
0=\mu(\l_l-\l_l^{-1})+\mathrm{\Lambda}\l_{\texttt{ts}}\l_l(\l_{\texttt{ts}}\l_l^2-1)\end{array}\right.  %\label{Eqts}
\end{align*}
and the nonlinear algebraic equation
\begin{equation*}
s_{\texttt{hs}}=\mu(\lambda_{\texttt{hs}}-\lambda_{\texttt{hs}}^{-1})+\mathrm{\Lambda}\lambda_{\texttt{hs}}^2(\lambda_{\texttt{hs}}^3-1),%\label{Eqhs}
\end{equation*}
respectively. 

Finally, we note that the Neumann boundary condition (\ref{BVP-z-theory})$_2$ may be replaced by the Dirichlet boundary condition $z_k=0$ at the front of any pre-existing crack and sharp corner that the boundary $\partial\mathrm{\Omega}_0$ of the specimen may feature. For the trousers test of interest in this work, we set $z_k=0$ as the boundary condition at the front of the pre-existing crack.

\section{Phase-field simulations of the trousers test for an elastic brittle material with Neo-Hookean elasticity}\label{Sec: Simulations Non-Linear}

Having introduced the fracture problem of interest in Section \ref{Sec: Trousers} and having recalled the phase-field approach to fracture in Section \ref{Sec: Phase-field Theory}, we are now ready to deploy the phase-field model  (\ref{BVP-y-theory})-(\ref{BVP-z-theory}) to simulate the canonical Neo-Hookean trousers test and to compare its predictions with the Griffith results presented in Subsection \ref{FEvsGRT}. 

\paragraph{Specimen geometry} Exactly as in Subsection \ref{FEvsGRT}, we consider a specimen of length $L=100$ mm and height $H=40$ mm in the $\bfe_3$ and $\bfe_1$ directions and constant thickness $B=1$ mm in the $\bfe_2$ direction. The specimen contains a pre-existing edge crack of length $A=50$ mm in the $\bfe_3$ direction; see Fig.~\ref{Fig1}. 

\paragraph{Material constants} The specimen is taken to be made of an elastic brittle material with the same Neo-Hookean elasticity constants used above, to wit,
\begin{equation*}
\mu=0.5\,{\rm MPa}\quad {\rm and}\quad \mathrm{\Lambda}=1\,{\rm MPa},
\end{equation*}
uniaxial tensile and hydrostatic strengths
\begin{equation*}
\sts=0.3\,{\rm MPa}\quad {\rm and}\quad \shs=1\,{\rm MPa},
\end{equation*}
and toughness
\begin{equation*}
G_c=200\,{\rm N}/{\rm m};
\end{equation*}
like that for $\mu$, the specific value for $G_c$ used here is representative of the SBR elastomers studied by \cite{Greensmith55}. Note that for these material constants, $W_{\texttt{ts}}=42.31$ kPa and $W_{\texttt{hs}}=229.19$ kPa.

\paragraph{Computational aspects} Given the height $H=40$ mm of the specimen and given the heuristic bound $3G_c/$ $(16W_{\texttt{ts}})\leq 0.89\,{\rm mm}$ on the material length scale associated with uniaxial tension, regularization lengths that are small enough to lead to results representative of sharp cracks are expected to satisfy the bound $\varepsilon<0.89\,{\rm mm}$. The results presented below correspond to the three values
\begin{equation}
\varepsilon=0.2, 0.3, 0.45\,{\rm mm},\label{eps-values}
\end{equation}
which permit to illustrate the convergence as $\varepsilon\searrow 0$.

\begin{figure}[H]\centering
 \includegraphics[width=3in]{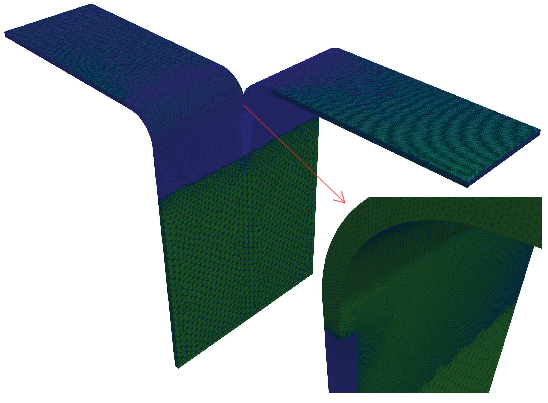}
\caption{\small One of the FE meshes used to simulate the trousers test showing a close-up around the initial crack front.}
   \label{Fig7}
\end{figure}

Given the thickness $B=1$ mm of the specimen, in view also of the values (\ref{eps-values}) of the regularization length, we make use of a FE discretization that around and ahead of the pre-existing crack --- where fracture propagation is expected to occur --- is of size
\begin{equation*}
\texttt{h}=0.05\,{\rm mm}.
\end{equation*}
Away from such a region of refinement, the FE discretization is coarser. Figure \ref{Fig7} shows one of the meshes used to carry out the simulations.

\begin{figure}[b!]\centering
 \includegraphics[width=3.3in]{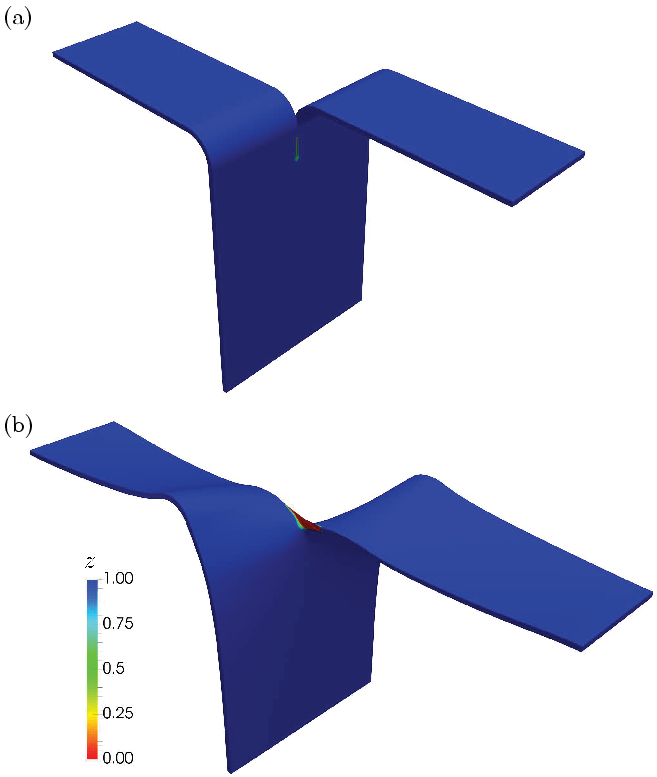}
\caption{\small Contour plots of the phase field $z(\bfX,t)$ over (a) the undeformed and (b) the deformed configurations at the applied deformation $l(t)=112$ mm between the grips, as predicted by the phase-field model (\ref{BVP-y-theory})-(\ref{BVP-z-theory}) for $\varepsilon=0.2$ mm.}
   \label{Fig8}
\end{figure}

\paragraph{Results and conclusions} Figure \ref{Fig8} presents representative contour plots of the phase field $z(\bfX,t)$ predicted by the phase-field model (\ref{BVP-y-theory})-(\ref{BVP-z-theory}) for the trousers test. The results, which are shown over the undeformed and the deformed configurations and pertain to the applied deformation $l(t)=112$ mm between the grips, illustrate that the specimen deforms and fractures in a self-similar manner. 

\begin{figure}[t!]
  \subfigure[]{
   \label{fig:8a}
   \begin{minipage}[]{0.5\textwidth}
   \centering \includegraphics[width=2.8in]{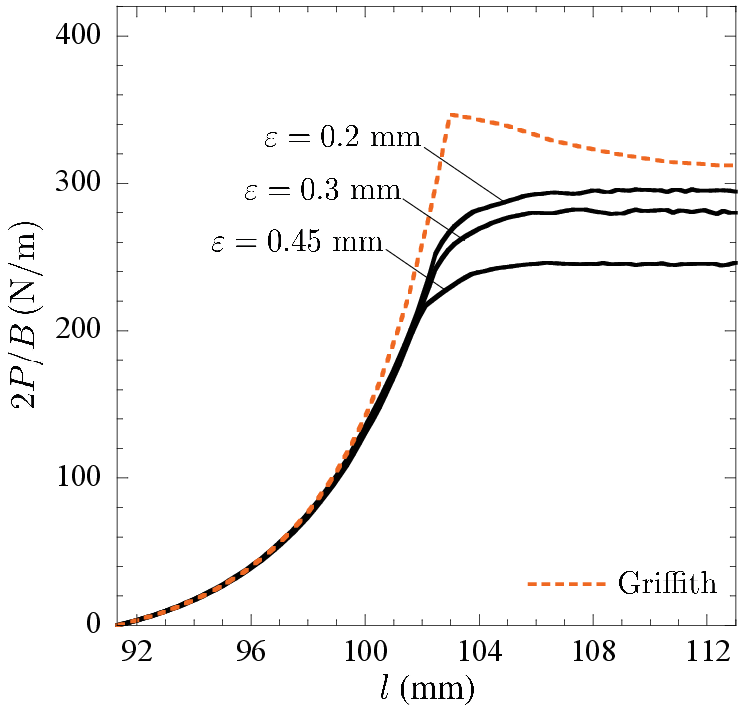}
   \vspace{0.1cm}
   \end{minipage}}
  \subfigure[]{
   \label{fig:8b}
   \begin{minipage}[]{0.5\textwidth}
   \centering \includegraphics[width=2.8in]{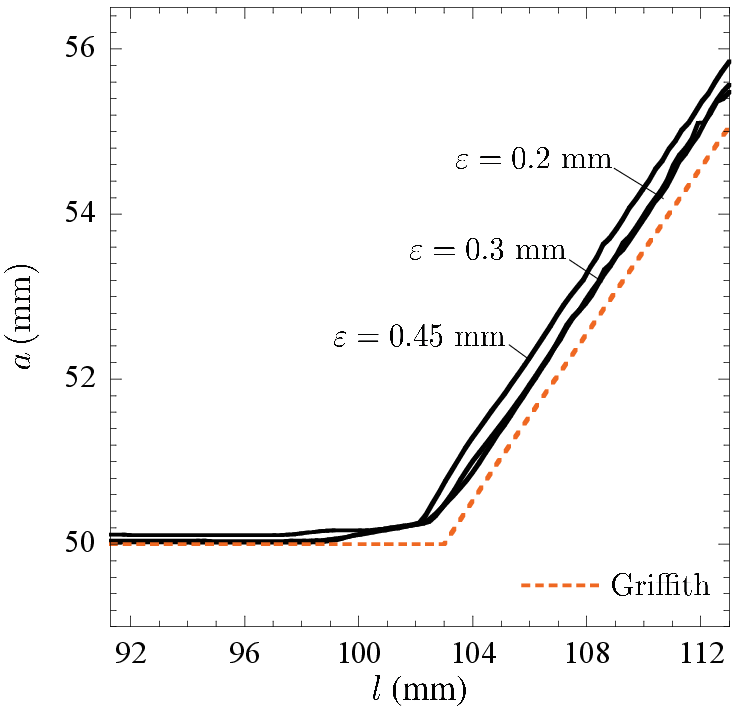}
   \vspace{0.1cm}
   \end{minipage}}
   \caption{Comparisons between the predictions generated by the phase-field model (\ref{BVP-y-theory})-(\ref{BVP-z-theory}) and the corresponding Griffith results. (a) The normalized force $2P(t)/B$ at the grips and (b) the crack length $a(t)$ as functions of the applied deformation $l(t)$ between the grips. The phase-field results are shown for three different values of the regularization length $\varepsilon$.}\label{Fig9}
\end{figure}

Figure \ref{Fig9} compares the normalized force $2P(t)/B$ and crack length $a(t)$ predicted by the phase-field model (\ref{BVP-y-theory})-(\ref{BVP-z-theory}) with the corresponding results that emerge from the Griffith energy criterion (\ref{Gc-0}), as computed from full-field FE solutions in Subsection \ref{FEvsGRT} above. Both sets of results are presented as functions of the applied deformation $l(t)$ between the grips. 
%
%%%%%%%%%%%%%%%%%%%%%%%%%%%%%%%%%%%%%%%%%%%%%%%%%%%%%%%%%%%%%%%%%%%%%%%%%%%%%%
%%%%%%%%%%%%%%%%%%%%%%%%%%%%%%%%%%%%%%%%%%%%%%%%%%%%%%%%%%%%%%%%%%%%%%%%%%%%%%
%

According to the Griffith energy criterion (\ref{Gc-0}), the crack starts to propagate when the force $P(t)$ at the grips reaches the value $P_c=0.174$ N, so that $2P_c/B=348$ N/m. The corresponding critical value for the deformation $l(t)$ between the grips is $l_c=102.98$ mm. As $l(t)$ increases beyond $l_c=102.98$ mm, the crack continues to propagate and the force $P(t)$ at the grips decreases slightly from $P_c=0.174$ N towards a lower value. Figure \ref{Fig9}(a) shows that the phase-field predictions are in agreement with these results for sufficiently small values of the regularization length $\varepsilon$.

Furthermore, according to the Griffith energy criterion (\ref{Gc-0}), the crack length remains put at $a(t)=A=50$ mm up until the time at which the force $P(t)$ at the grips reaches the critical value $P_c=0.174$ N, after which point it grows as $a(t)=A +f(l(t))$, where the function $f(l(t))$ is approximately --- but \emph{not} exactly --- $f(l(t))=(l(t)-l_c)/2$. Similar to those in Fig.~\ref{Fig9}(a), the comparisons in Fig.~\ref{Fig9}(b) show that the phase-field predictions are in agreement with this result for sufficiently small values of the regularization length $\varepsilon$.

Summing up, the results in Figs.~\ref{Fig8} and \ref{Fig9} show that the phase-field approach to fracture --- in particular, the phase-field formulation introduced by \cite*{KFLP18} --- describes Mode III fracture propagation in a manner that is consistent with the Griffith  competition between the bulk deformation energy and the surface fracture energy.

We close by remarking that the trousers test studied here is ideal to probe the capabilities of computational models --- not just phase-field models --- to describe fracture propagation. Insomuch that it should be one of the challenge problems to pass in any problem set envisioned to establish the viability of fracture models.

\section*{Acknowledgements}

This work was supported by the 3M Company and the National Science Foundation through the Grant DMS--2308169. This support is gratefully acknowledged.

\bibliographystyle{unsrtnat}
\bibliography{References}

\end{document}